\documentstyle{article}

\author{Ramil' F.Bikbaev, Vadim R.Kudashev\\
Institute of Mathematics, Chernyshevskii 112, Ufa, 450000, Russia\\
E-mail: vadkud@nkc.bashkiria.su}
\date{February 24, 1994}

\title{Example of shock wave in unstaible medium: The focusing
nonlinear Schrodinger equation }

\begin{document}

\maketitle
\begin{abstract}
Dissipation\-less shock waves in modulational unstable one-dimensio\-nal
medium are investigated on the simplest example of integrable focusing
nonlinear Schr\''odin\-ger (NS) equation. Our approach is based on the
construction of special exact solution of the Whitham-NS system, which
''partially saturates'' the modulational instability.
\end{abstract}

\section{Introduction}

It is well known that focusing NS equation
\begin{equation}
iu_t+u_{xx}+2\mid u\mid ^2u=0,
\end{equation}
may be rewritten as hydrodynamic type system
\begin{equation}
\begin{array}{cc}
f_t+2(f\cdot v)_x=0, & v_t+2vv_x-2f_x=\left( 2f_{xx}/f-f_x^2/f^2\right)
_x/4,
\end{array}
\end{equation}
where
\begin{equation}
\begin{array}{cc}
u=\sqrt{f}\cdot \exp (i\varphi ), & v=\varphi _x.
\end{array}
\end{equation}
We consider (1)-(3) with the following step-like initial condition
\begin{equation}
\sqrt{f(x,0)}=\left\{
\begin{array}{c}
2,x<0 \\
1,x\geq 0
\end{array}
\right. \sim \left\{
\begin{array}{c}
2a,x<0 \\
a,x\geq 0
\end{array}
,\right. v(x,0)=0.
\end{equation}
It is natural to call the solution $[f(x,t),v(x,t)]$ or $u(x,t)$ of the
problem (1)-(4) the {\it shock wave solution}.

The main purpose of this Letter is to describe the qualitative behavior of
the shock wave at $t\gg 1$. We note that this problem is much more difficult
than the analogous problem for stable models [1,2].

The typical and simplest shock problem in modulational stable model was
first studied by Gurevich and Pitaevskii [1] on the example of Korteveg-de
Vries (KdV) equation. In [1] modulational theory developed by Whitham [3]
was used. It is important that the Whitham system for KdV equation as well
as for other stable models [3] is hyperbolic, which means that Whitham
characteristic speeds are everywhere real: $Im(S_i)\equiv 0.$ Howere for
unstable models these speeds are generally speaking complex [3]: $%
Im(S_i)\neq 0$ (elliptic case). This complicates very much any analytical
investigation of unstable models due to exponential growth of linear
perturbation of the (locally) constant background. In particular there are
obvious obstacles in studying of the shock problems in unstable systems,
which are closely related to the classical problem of the modulational
instability of monochromatic wave in (1) (see for example [4]).

In this Letter we demonstrate how to overcome these difficulties on the
concrete example (1)-(4). We shall show that famous strategy [1] may be
modified and successfully applied to (1)-(4). Our approach is based on the
conception of ''partially saturating modulational instability solutions'' of
elliptic Whitham-NS equations proposed in [5] for description of
modulational instability in (1). The key role in our construction plays a
special subclass of Whitham-NS system solutions with vanishing imaginary
parts $Im(S_i)=0$ of some (not all!) characteristic speeds (see (12), (17)).

\section{One-phase NS equation solution and its Whi\-tham-NS modulations.}

It is well known that one-phase solution of (1), (2) has form
\begin{equation}
\begin{array}{cc}
f_I(\theta )=f_3+(f_1-f_3)dn^2\{\sqrt{f_1-f_3}\cdot \theta ;m\}, &
m=(f_1-f_2)/(f_1-f_3),
\end{array}
\end{equation}
\begin{equation}
\begin{array}{cc}
v_I(\theta )=U/2-A/f(\theta ); & \theta \equiv x-Ut,
\end{array}
\end{equation}
where $f_1\geq f\geq f_2\geq 0\geq f_3,A=\sqrt{-f_1f_2f_3}\geq 0,dn$ - is
the Jacobi elliptic function. Elliptic spectral curve corresponding to (5),
(6) has branching points $\lambda _i$, $i=1,2,3,4;$ $\lambda _2=\lambda
_1^{*},\lambda _4=\lambda _3^{*}$ such that (c.f. [6])
\begin{equation}
\begin{array}{c}
\lambda _1\equiv \alpha -i\gamma =U/4-
\sqrt{-f_3}/2-i(\sqrt{f_1}+\sqrt{f_2})/2, \\ \lambda _3\equiv \beta -i\delta
=U/4+\sqrt{-f_3}/2-i(\sqrt{f_1}-\sqrt{f_2})/2.
\end{array}
\end{equation}
Whitham-NS equations for (1), (2), (5)-(7) can be presented in diagonal form
(c.f. [6])
\begin{equation}
\begin{array}{cc}
d\lambda _i/dt+S_i(\lambda )d\lambda _i/dx=0, & i=1,2,3,4,
\end{array}
\end{equation}
\begin{equation}
\begin{array}{cc}
S_1=U+2\lambda _{12}/(1-\mu \lambda _{32}/\lambda _{31}), & S_3=U+2\lambda
_{34}/(1-\mu \lambda _{14}/\lambda _{13}), \\
S_2=S_1^{*},S_4=S_3^{*}; & \mu \equiv E(m)/K(m),
\end{array}
\end{equation}
where $\lambda _{ij}\equiv \lambda _i-\lambda _j$, $K,E$ are the complete
elliptic integral of the first and second kind respectively, $m=\lambda
_{21}\lambda _{43}/\lambda _{32}\lambda _{14}.$

\section{Solution of Whitham-NS system and long-time behavior of the
NS-hydro\-dynamic shock wave.}

We construct an approximate solution of the shock problem (1)-(4) for $t\gg
1 $ as follows. In the ''external'' region on $(x,t)$ plane: $%
(x<0,x>x^{+}(t))$, the solution is zero-phase one:
\begin{equation}
\sqrt{f_0(x,t)}=\left\{
\begin{array}{c}
2,x<0 \\
1,x\geq x^{+}(t)
\end{array}
,\right. v_0(x,0)=0.
\end{equation}

In the oscillation region $0\leq x\leq x^{+}(t)$ the ''internal'' solution
is given by the one-phase solution (5), (6) with modulated due to (8), (9)
parameters. In this region we use special solution of the Whitham-NS system
(8), (9):
\begin{equation}
\lambda _1\equiv const.,
\end{equation}
\begin{equation}
\begin{array}{cc}
Im(S_3)=0, & t\cdot Re(S_3)-x=g(\beta ,\delta ),
\end{array}
\end{equation}
where $g(\beta ,\delta )$ is an arbitrary smooth function of its variables,
which is determined from the initial conditions.

Let us consider the simplest case $g\equiv 0$ (c.f. [1], [5,7]). From the
initial condition (4) we get
\begin{equation}
\begin{array}{cccc}
\gamma =1, & \alpha =0, & \Longleftrightarrow & \lambda _1=-i.
\end{array}
\end{equation}

{\it Proposition.} The system (12), (13) with $g\equiv 0$ is compatible and
has unique solution with $\delta \geq 0,$ $\beta \geq 0$ in the ''internal''
region $0\leq x\leq x^{+}(t).$ Near the boundary $x^{+}=x^{+}(t)$ the
solution of the system (12) has the form
\begin{equation}
\begin{array}{ccc}
x^{+}=4\sqrt{2}t, & x=x^{+}-x^{\prime }, & 0<x^{\prime }\ll 1, \\
\beta \approx 1/\sqrt{2}-7x^{\prime }/48t, & \delta ^2\approx x^{\prime }/2
\sqrt{2}t. &
\end{array}
\end{equation}

At the boundary $x=x^{+}(t)$ the solution $(f_I,v_I)$ from (5), (6) is
continuously glued with $1$ from (4). If $(x/t)\rightarrow +0$ the points $%
(\lambda _3,\lambda _4)$ closely come to the points $(\lambda _1,\lambda _2)$%
. In this limit our solution $(f_I,v_I)$ from (5), (6) degenerates into the
stationary soliton (breather), corresponding to $\lambda _2=\lambda _4=i,$ $%
\lambda _1=\lambda _3=-i.$

{\it Remark 1.} Due to Galilean invariance of (1) the above analysis may be
easily extended to the case of the shock wave with more general initial
conditions
\begin{equation}
\sqrt{f(x,0)}=\left\{
\begin{array}{c}
2,x<0 \\
1,x\geq 0
\end{array}
,\right. v(x,0)=2b.
\end{equation}
The corresponding changes in the formulae (11)-(14) are as follows: $
\begin{array}{ccl}
x^{+}\rightarrow 4bt+x^{+}, & \lambda _1\rightarrow b+\lambda _{1,} &
\lambda _3\rightarrow b+\lambda _3.
\end{array}
$

{\it Remark 2. }The mirror symmetry $x\rightarrow -x$ allows to solve the
shock problem symmetric to (4)
\begin{equation}
\sqrt{f(x,0)}=\left\{
\begin{array}{c}
1,x\leq 0 \\
2,x>0
\end{array}
,\right. v(x,0)=0.
\end{equation}
We do not go into the details, but just note that corresponding solution of
Whitham-NS system (8) has the form
\begin{equation}
\begin{array}{ccc}
\lambda _3\equiv const., & Im(S_1)=0, & t\cdot Re(S_1)-x=0,
\end{array}
\end{equation}
and the picture of the solution $u(x,t)$ at $t\gg 1$ is symmetric with
respect to $x\rightarrow -x$ transformation to the above picture.

{\it Remark 3.} The problems (4), (15), (16) are of course the simplest
examples of the shock problems in unstable model (1), (2). As we have shown
above in these simplest cases the traditional ideology of [1,2] can be
applied (after proper modification). For more general initial data it is
necessary to invent new technique. For example, how to solve the following
shock problem%
$$
u(x,0)=\left\{
\begin{array}{c}
1,x<0 \\
0,x\geq 0
\end{array}
\right. ?
$$
We hope to answer this question in the nearest future.

{\it Remark 4.} Note that the above shock wave process (1)-(14) has the same
typical time $t\approx 1$ as ordinary modulational instability $(\mid
Im(S_1)\mid \cong 1$ as $t=0).$

{\bf Acknowledgement}

The work of V.R.K. was supported, in part, by ''BUT-RUTEX'', and by INTAS
grant.

{\bf References}

[1] A.V.Gurevich and L.P.Pitaevskii, JETP 38 (1974) 291.

[2] A.V.Gurevich and A.L.Krylov, Zh. Eksp. Teor. Fiz. 92 (1987 ) 1684;
R.F.Bikbaev, Teor. Mat. Fiz. 81 (1989) 3; R.F.Bikbaev, Zap. nauch. sem.
POMI, 199 (1992) 25.

[3] G.B.Whitham, Proc. Roy. Soc. A283 (1965) 238; Linear and nonlinear
waves, John Wiley and sons, 1974, N.Y.

[4] B.B.Kadomtsev, Collective phenomena in plasmas, Moscow, Nauka, 1976 [in
Russian].

[5] R.F.Bikbaev and V.R.Kudashev, Whitham deformations partially saturating
the modulational instability in the nonlinear Schrodinger equation, preprint
Los Alamos Nonlinear Science PBB: patt-sol/9404003, JETP Lett. (1994),
submitted.

[6] M.V.Pavlov, Teor Mat. Fiz. 71 (1987) 351.

[7] A.M.Kamchatnov, Phys. Lett. 162A (1992) 389.

\end{document}